\documentclass[11pt]{article}

\usepackage{graphicx}
\usepackage{amssymb}
\usepackage{amsmath}

\begin{document}

\title{What Makes a Computation Unconventional?\\  or, there is no 
such thing as Non-Turing Computation}

\author{S. Barry Cooper\\ University of Leeds, UK}
\date{}

\maketitle

\enlargethispage{12pt}
Turing's standard model of computation, and its 
physical counterpart, has given rise to a  
powerful paradigm. There are assumptions 
underlying the paradigm which constrain our thinking about 
the realities of computing,  
not least when we  doubt the paradigm's adequacy. 

There are assumptions concerning the logical 
structure of computation, and the character of its 
reliance on the data 
it feeds on. There is a corresponding complacency 
spanning theoretical --  
but not experimental -- thinking about the  complexity of 
information, and its mathematics. We point to 
ways in which classical computability can 
clarify the nature of apparently unconventional 
computation. At the same time, we seek to expose 
the devices used in both theory and practice 
to try and extend the scope of the standard model. 
This involves a drawing together of different 
approaches, in a way that validates the intuitions 
of those who question the standard model, 
while providing them with a 
unifying vision of diverse routes ``beyond 
the Turing barrier". 

The results of such an analysis 
are radical in their consequences, and break the mould 
in a way that has not been possible previously. The aim 
is not to question, invalidate or supplant the richness 
of contemporary thinking about computation. 
A modern computer is not {\em just} a universal 
Turing machine. But the understanding the model 
brought us  was 
basic to the building of today's digital age. It gave us 
{\em computability}, an empowering insight, and 
computing with {\em consciousness}. 
What is there fundamental that  unconventional computation directs us to?  
What {\it is it} makes a computation unconventional? 
And having fixed on a plausible answer to this question, we ask: 
To what extent can the   explanatory power of the mathematics 
clarify key issues relating to emergence, 
basic physics, and the supervenience of 
mentality on its material host?

\section{METHOD OVER MATTER}

There is a huge  literature concerning computability. 
It has grown beyond what anyone might have anticipated 
back in the 1930s. 
The subject has taken on a life of its 
own, the context has spread across disciplinary boundaries in a startling way and 
ideas are increasingly hard to categorise and evaluate within traditional structures. 
During the  2012 centenary 
of Alan Turing's birth, we were reminded that some of the most important early contributions to 
the computing revolution came from people who thought deeply about the way the 
world computes, while gaining strength from their independence of what other people had 
done. Thank you, Turing.  Relieved of the burden of the taxonomy of who said 
what and when, let us start by mapping out some of the main features of the 
standard model of computation. Or rather, the main features of how 
a starting student in computability might unwrap the model. \enlargethispage{12pt}

The most basic feature, a feature not just of Turing's 1936 Turing machine but of 
all the equivalent models, is 
its {\em disembodiment}. We might have been told at some point that 
it was devised as a disembodying model of machine computation. Not so, of course. 
2012 has made everyone aware of the very specific 
physicality of the computing situation that Turing was modelling, 
the predominantly women `computers'  following instructions. 
One of the special strengths of the Turing model is its close 
relationship with physical computation, via a very specific 
deconstruction of a {\em typical} computational context. We 
have but a confused idea of what a machine might be. We have 
a firmer grip on what a `computer' following instructions might 
be doing, using well-defined workspace, tools and conventions. 
The underlying physicality may be highly complex. But such things 
as the human computer's aches and pains, her feelings of hunger 
or boredom, are factored out of the process. We extract 
emergent features of the material context which are far from 
disembodiments of the computation, but which give us a model 
which we may re-embody in quite different contexts, and whose 
mathematical properties can be investigated with a realistic 
hope of relevance to a wide spectrum of controlled 
situations. 

This provides a green light to those who would turn such a 
superficial take on physical computation back on its host.  In the 
absence of a corresponding deconstruction of more complex 
physicality -- and ignoring the fact that even more ad-hoc descriptions 
of particle physics, life, cosmology, human mentality etc. are incomplete -- 
the temptation is to turn the Turing model into metaphor, and then into 
extended model, in ways we can only argue about. 
The disembodiment implicit in the standard model is not 
so simple. It has a character which we should not 
ignore. Surfing the computational world is fun, but the underlying 
complexity may still surprise. 

We have dwelled on the basic particularity and oddness of the standard model. 
It is relevant to what we expect of other aspects of the dominant paradigm. 
Before Turing and logicians like Emil Post, Kurt G\"odel, Alonzo Church, and 
Stephen Kleene came on the scene, an important input to a computation 
was the computing machine itself. Physically the machine embodied a weighty piece 
of data. The logician's overview provided an extraction of its essentials in the form 
of a simple code -- a natural number, a finite binary string, or other similar 
mathematical object. This provided two hugely important features 
of the standard model, and the modern computational world. 

Firstly, however we structure our machines, their descriptions 
can be converted into data used by machines based on a different 
logical analysis, enabling the construction of algorithms 
for converting `programs' within one framework into one 
fitting with another. Much activity of the early investigators 
involved devising such algorithms (say, for converting 
a description of a $\lambda$-computable function 
into one for an equivalent Turing computable function). 
The natural conclusion was that Turing machines could 
`compute anything'. 

Secondly, having trivialised the hardware, 
the power of the 
computing paradigm lay in the programming. 
And with a model which turned machines into data, 
it was a short step to building a machine which could 
mine machines for different purposes  out of 
whatever data you gave it -- and having decoded the 
program from the data, could implement it. This was the origin of  
the so-called `universal Turing machine'. By a simple mathematical 
sleight of hand, one had moved machine from physical world 
to the realm of pure thought. Well, not quite, as Charles Babbage 
found out in the process of getting his Analytical Engine built 
prior to 1871. Of course, Babbage's machine, like others 
pre-dating the 1948 Manchester Small-Scale Experimental Machine (SSEM), 
or `Manchester Baby', was not universal machine in the sense of a 
modern stored-program computer.

Today's computers are a true embodiment of Turing's 
universal machine in that they enable programs to be combined and edited in 
increasingly creative ways, without the need for any rebuilding of the 
programmer-computer physical interface. Nowadays, the program, once input, 
becomes part of the computer, to be stored, adapted or discarded by the 
programmer 
without any grappling with punched cards or realigning of wired connections 
or switches. Early designs of Babbage, Konrad Zuse and others are 
`Turing complete', but lack the vital stored-program feature. This 
important ingredient of Turing's 1936 logical analysis was incorporated  
by John von Neumann in his June 1945 EDVAC report, and also features,  
much less influentially, in Turing's report on the ACE of later that same year. 

Universality; the transposability of computational activity from one computing platform 
to another; the supplanting of the physical by the logical; the redundancy of 
information beyond the type 1 or type 2 mathematical level --- these are 
familiar aspects of an overarching computational paradigm. The underlying 
assumptions have served us well, and moulded our thinking about the wider 
context. One can recognise it in early approaches to artificial intelligence. 
In the philosophy of mind one has various functionalist viewpoints, 
with Hilary Putnam explicitly drawing on the universal machine metaphor 
in his seminal 1960s writings on the topic. 
Again, in computer science one has the allied notion of `virtual machine' 
quite validly useful in both practical and theoretical contexts. 
One observes the paradigm in the drive to reduce social interaction and development to 
the algorithmic, setting complex interactive processes within 
simple rule-based game structures. The feedback between the emergent and the 
algorithmic, to which we return below, does not fit well with  
`corporate thinking', with its drive for strategic certainty. 

In computer science and mathematics the paradigm 
can be detected in sophisticated approaches 
to the logic of computation, focused on the value of frameworks transferable 
not just between specific contexts, but between different disciplines. 
Categorical methods have been productive in the computational 
context, where according to Samson Abramsky 
``in the work on 
concurrent processes, the behaviour {\em is} the object of interest. 

\section{PROCESS AND EMBODIMENT}

History brings its own reminders that computers are not `just' universal 
Turing machines. Moving the model from the human `computers' 
platform to a more efficient and cost-effective electronic platform was 
highly non-trivial. Apart from this, the re-embodiment of 
computing brought us closer to the main point of Turing's 1936 
paper --- a proof that there are interesting questions beyond 
the reach of algorithms. In retrospect, one can rephrase this as 
``computers are stupid", and go on to ask if a 14 billion year old universe 
is subject to similar limitations. Is the mathematics of Turing's 
simple diagonalisation of the computable reals unembodiable? 

The sort of problems Turing and Alonzo Church showed to be 
unsolvable by a computer were very natural in an everyday sense. 
From Turing we know that if $\mathbf U$ is a universal Turing machine 
then there is no computer can tell us , for an arbitrary input $ x$, 
whether  $\mathbf U$ will ever produce an output from $ x$. 
This is the `unsolvability of the Halting Problem' 
for $\mathbf U$, with the set of numbers $x$ on which $\mathbf U$ 
halts called the Halting Set for $\mathbf U$. 
Remembering that $x$ can code a program, this gives us an indication 
of why computer program checking is such a tough problem. 
The process tends to be experimental, with a new piece of 
software requiring a sequence of updates to fix various bugs. 

Even closer to home  is `Church's Theorem' --- actually Turing 
proved it too, it is just the negative solution to Hilbert's 
{\em Entscheidungsproblem}. This says that if you have a 
sentence in everyday language (as formalised in first-order 
logic), there is no computer that can tell us of any such sentence 
whether it is logically valid or not. To many this is quite informative 
and counter-intuitive.

One can extract from each of these problems a binary expression 
for a real number $r$. Say $r=0\cdot r_0r_1 \dots $, with each $r_i = 0$ or $1$, 
where $r_i=1$ exactly when $\mathbf U$ successfully computes 
on input $x=i$. $\mathbf U$ can be thought to `compute' $r$ in the 
sense that this number is uniquely decided by the actions of $\mathbf U$ 
in computing on $0,1,2, \dots$ successively. $r$ is a very real {\em feature} 
of the real world in which $\mathbf U$ lives and operates. However, 
the level of abstraction of $r$ means that even though we can `see' 
$\mathbf U$ computing, we cannot `see' $r$ at all well. If we could see 
$r$ we would be moved to allow that it it is `computed' in some sense. 
Of course, $r$ does not fit into the classical paradigm, since 
$r$ is not available as an input to further computation by $\mathbf U$. 
Not only can {\em we } not see $r$, nor can $\mathbf U$. And this is 
not just due to the incomputability. $r$ as a mathematical object is 
of higher {\em type} than the natural numbers which $\mathbf U$ usually 
accepts as an input. Anyway, in the absence of an embodied presence, 
$r$ is not considered a computed outcome. 

Mathematically, $r$ is {\em definable} from $\mathbf U$, but the 
existential quantifier needed to define it puts $r$ on the other side 
of an unembodied chasm. 
The question arises: Can this chasm be crossed given the 
right material conditions?

There is an obvious counterpart of this elevation of type, and 
Turing's proof of a resulting incomputability. The non-locality 
of the view amounts to a logical interactivity between computations. 
The observing process provides the connectivity, with us a player 
in the physical environment. We are no longer in the presence of 
an individual computation, it is an interactive {\em process} at work, 
with what we will subsequently recognise as an {\em emergent} 
incomputable real $r$. We will come to regard {\em emergent} as 
real world analogue of `definable'. Emergence plays an important 
role in many dynamical systems, such as weather systems, 
large scale social interactions, the internet, biology, creative 
thinking, and turbulent environments of many kinds. 

Definability is commonly ignored, or regarded 
as a logician's playground, with important 
instantiations in the wider mathematical context. 
A useful `missing link' is the fractal, with both 
precise mathematical description and a visual presence, 
often enhanced via computer simulation. 

In a formal sense, the Halting Problem is in the same world as 
the Mandelbrot set, for instance. We have gone up another level of the 
type structure, but there is an in-principle connection. We have a simple 
computable rule hosted by the complex numbers. Based on this there 
is a two-quantifier definition of the members of the Mandelbrot set, 
which with a little manipulation can be reduced to a one-quantifier 
expression for the purpose of the well-known computer simulation. 
The computability or otherwise  of the Mandelbrot set is still an open 
problem. But unlike our incomputable Halting Set, the 
Mandelbrot set comes beautifully and interestingly embodied, 
with quite visual counterparts to the suspected incomputability. 
There it is on our computer screen. And we can delve as deeply 
as we like into this 
fascinatingly surprising type-2 object. The reason for this is that we 
are {\em sampling} this set of complex numbers. The computer 
screen image involves a trick reduction of type. Turing himself 
was familiar with the usefulness of statistical sampling for 
reducing complex information to something computationally 
approachable. It is not a purely ad hoc methodology. It is a way 
of recognising the higher type computability enabled by a 
definition, or by some real world process to whose 
computational underpinnings we are not privy. 

Moving beyond our mathematical comfort zone, we can observe 
many everyday examples of emergence as instances of objects 
definable in a real context. We see apparently chaotic environments 
involving generation of informational complexity via simple rules 
with a computational character. And we further observe the 
accompanying, often surprising, emergence of new 
regularities --- such as those of Robert Shaw's dripping taps --- 
entropic resting points, often at most observable via the sort 
of selective sampling which made visible the embodied 
Mandelbrot set. 

The embodied computation of higher type objects is not in itself 
a challenge to the classical model. But its character does 
mell with intuitions concerning unconventionality of computation. 
And the parallel with the established incomputable $r$ 
and its mathematical context certainly rings the alarm bells.

\section{EMERGENCE AND 
DEFINABILITY}

All around us we see a world exhibiting 
algorithmic content accompanied by 
hierarchical structure not easily explainable in 
terms of the familiar underlying rules. 
Our everyday lives are built around 
what appears to be a computable 
environment, but nature continually 
surprises us. Much of that surprise 
is attached to natural form which does 
appear to be part of a universe which 
`knows what it is doing', and it is this 
we think of as `emergent'. 

The importance 
of getting a mathematical grip on this 
omnipresent phenomenon --- in evidence from 
`strange attractors' to human creativity, and 
from the origins of life to large-scale cosmic structure --- 
is illustrated by the history of `British Emergentism',  
and its heyday in the 1920s. One of the leaders 
of the movement was the Cambridge philosopher 
C. D. Broad. Here he is\footnote{C.D. Broad, {\em The Mind 
and Its Place In Nature}, Kegan-Paul, London, 1925, p.59.} 
  in 1925, attempting an explanation 
of what emergence is, while pointing to 
illustrative examples:
\begin{quote}
$\dots$ the characteristic behaviour of the whole $\dots$ 
could not, even in theory, 
be deduced from the most complete knowledge of the behaviour 
of its components $\dots$ This .$\dots$ is what I understand by the 
`Theory of Emergence'. I cannot give a conclusive example of it, since 
it is a matter of controversy whether it actually applies to anything $\dots$ 
I will merely remark that, so far as I know at present, the characteristic 
behaviour of Common Salt cannot be deduced from the most complete 
knowledge of the properties of Sodium in isolation; or of Chlorine in 
isolation; or of other compounds of Sodium, $\dots$
\end{quote}
Dramatic scientific developments were in progress around this 
time. 1925 saw the key elements of the new quantum mechanics 
put in place by Werner Heisenberg and Erwin 
Schr\"odinger, and by the 5th Solvay conference in 1927 
quantum theory was revolutionising the  foundations of chemistry.  
The mystery was stripped from the examples from chemistry of Broad 
and others.\footnote{ See Brian McLaughlin's  
 article ``The Rise and Fall of British Emergentism", in 
 {\em Emergence or Reduction? -- Essays on the Prospects of Nonreductive Physicalism} 
 (A. Beckermann, H. Flohr, J. Kim, eds.), de Gruyter, Berlin, 1992, pp.49--93.} 
 
 For Stuart Kauffman\footnote{Stuart Kauffman, 
{\em Reinventing the Sacred: A New View of Science, Reason and Religion}, Basic Books, 2008, p.281.}  
 emergence is not just an example of unconventional 
 computation, it calls into question basic assumptions about the 
 computational 
 content of causality and the 
 deterministic character of the universe:
 \begin{quote}
 We are beyond reductionism: life, agency, meaning, value, 
 and even consciousness and morality almost certainly arose naturally, 
 and the evolution of the biosphere, economy, and human culture are 
 stunningly creative often in ways that cannot be foretold, indeed in 
 ways that appear to be partially lawless. The latter challenge to current science 
 is radical. It runs starkly counter to almost
four hundred years of belief that natural laws will be sufficient to explain 
 what is real anywhere in the universe, a view I have called the Galilean spell. 
 The new view of emergence and ceaseless creativity partially beyond 
 natural law is a truly new scientific worldview in
which science itself has limits.
 \end{quote}
 Such claims are counterbalanced by words of caution from 
 Ronald Arkin\footnote{Ronald C. Arkin, {\em Behaviour-Based Robotics}, MIT Press, 1998, p.105.}:
 \begin{quote}
 Emergence is often invoked in an almost mystical sense regarding the capabilities 
 of behavior-based systems. Emergent behavior implies a holistic capability 
 where the sum is considerably greater than its parts. It is true 
 that what occurs in a behavior-based system is often a surprise to the 
 system's designer, but does the surprise come because of a 
 shortcoming of the analysis of the constituent behavioral building 
 blocks and their coordination, or because of something else?
 \end{quote}
 
In the face of historic confusions, and radical contemporary 
speculations, the clarifying role of mathematics 
is urgently needed. This is not to brush aside the more detailed 
 proposals of Kauffman and others. The aim is to place 
 them in a more foundational framework. 

To this end, one needs more than the codifying of current 
`best observational practice' represented by 
the Test of Emergence of Ronald, Sipper and Capcarr\`ere in 
{\em Design, observation, surprise! A test of emergence} (Artificial Life, {\bf 5} (1999), 225--239). 
Here is a summary of their qualifying criteria:
\begin{enumerate}
\item[1)] {\bf Design}: The system has been constructed by the designer, 
by describing local elementary interactions between components (e.g., artificial 
creatures and elements of the environment) in a language $\mathfrak{L}_1$.
\item[2)] {\bf Observation}: The observer is fully aware of the design, but describes 
global behaviors and properties of the running system, over a period of time, using a 
language $\mathfrak{L}_2$.
\item[3)] {\bf Surprise}: The language of design $\mathfrak{L}_1$ and the language 
of observation $\mathfrak{L}_2$ are distinct, and the causal link between the 
elementary interactions programmed in $\mathfrak{L}_1$ and the behaviors 
observed in $\mathfrak{L}_2$ is non-obvious to the observer -- who therefore experiences 
surprise. In other words, there is a cognitive dissonance between 
the observer's mental image of the system's design stated in $\mathfrak{L}_1$ and 
his contemporaneous observation of the system's behavior stated in $\mathfrak{L}_2$.
\end{enumerate}
Might this  serve as a test for unconventional computation? Unconventionality 
certainly requires some obstacle to reduction to basic algorithmic 
structure. And it is hard to {\em design} a computational device which 
has no underpinning of classical ingredients. On the other hand, 
there are potentially incomputable processes in nature for which 1) or 2) fail. 
Can a foundational approach make computational sense of the 
outcome of a quantum measurement leading to a collapse of the 
wave function?

A nice aspect of the above test is its differentiation between `designer' and 
`observer' languages. This is a feature of the fragmentary nature of science, 
where it is common to view, say, biology as emergent from an underlying 
quantum mechanical base,  with its own emergent rules and language, 
non-reducible to the quantum level on which it depends. In the 
case of the Halting Set for a universal Turing machine, 
$\mathfrak{L}_2$ is distinguished by the addition of quantification. 
\vspace{4pt}

Alan Turing recognised something computationally interesting 
in emergence when he investigated the mathematics 
of morphogenesis. In the early 1950s Turing wrote his 
groundbreaking paper on  {\em The chemical basis of Morphogenesis}, 
in which he proposed a simple reaction-diffusion system describing 
chemical reactions and diffusion to account for morphogenesis in 
a range of cases. He even ran computer programs on the 
early Manchester Mark 1 computer (a more powerful successor 
to the `Baby') with the aim of verifying his reaction-diffusion 
`design' underlying such emergent patterns as the familiar 
black and white dappling on a Holstein dairy cow. 

What is specially interesting about this work is how it 
related the powerful descriptive framework of differential 
equations to emergent form in nature, so exhibiting a 
connection between the mathematics of higher type 
objects and apparent emergence. It is hard to claim 
computational unconventionality on this basis -- 
the solutions to Turing's equations tended to be 
computable -- but then mathematics provides 
us with little means of identifying real world 
incomputability. Reducing the Halting Problem 
to an elusive solution to a non-linear differential 
equation is not very likely. on the other hand, 
Marian Pour-el and Ian Richards had 
some success designing  
`A computable ordinary differential equation which possesses no 
computable solution'\footnote{In: Annals of Mathematical Logic
Volume 17, November 1979, Pages 61Ð90.}

To summarise: Turing provided examples of descriptions 
of emergent phenomena, whereby one might characterise 
the emergence as an expression of a higher type 
computation. And this fits well with the Ronald-Sipper-Capcarr\`ere test 
for emergence, via the provision of each of design, observation 
and surprise. With the latter mathematically traceable back 
to the type-climbing and concomitant potential 
incomputability of the emergent form. 
\vspace{4pt}

Is it pure serendipity the discovery that some phenomena 
can be described in terms of material context? There is a 
strong intuition that form in the universe arises for a reason. 
Scientifically this intuition takes the form of an 
expectation of finding descriptions of phenomena in terms 
of basic laws of nature. An echo of such an expectation be traced back to 
Gottfried Leibniz's 1714 description\footnote{See {\em The 
Monadology}, sections 31, 32.} of his `principle of sufficient 
reason':
\begin{quote}
\dots there can be found no fact that is true or existent, or any true 
proposition, without there being a \underline{sufficient reason} for its being so 
and not otherwise, although we cannot know these reasons in most cases.
\end{quote}

The intuition that natural phenomena not only 
generate descriptions, but arise and derive from them, 
connects with a useful abstraction associated with 
Alfred Tarski, and growing out of his 1930s work on the notion 
of truth for formal languages. {\em Mathematical definability}, 
or more generally {\em invariance} under 
automorphisms of an appropriate structure, provides 
an effective organiser of the relative ontology of relations 
over a structure. 

Definability is a basic notion which deserves to be better known in the mathematical 
world, and in the wider scientific community. It's relevance to physics 
has been long recognised. Hans Reichenbach worked to 
axiomatise Einstein's relativity in the 1920s, a project carried forward 
in relation to general relativity today by the Budapest group of Istv\'an N\'emeti and 
Hajnal Andr\'eka. This extension of the fundamental mathematics enables us to 
deal with a wider range of phenomena, taking us beyond the classical 
computational model. It gives precision to our experience of emergence as a 
potentially trans-algorithmic determinant of phenomena. 

The overarching aim now is to describe global relations in range of contexts in 
terms of local structure, so capturing the emergence of large scale formations. And mathematically 
to formalise such descriptions as definability, or as invariance 
over basic computational structure. Although Stephen Kleene provided 
formal content to the notion of higher type computation via a series of papers 
spanning over 30 years (1959--1991), the physical relevance of his take on the 
topic needs to be clarified. A forthcoming book on ``Computability At Higher Types" by 
John Longley and Dag Normann is eagerly anticipated. The intuition is that 
computational unconventionality certainly entails higher type computation, 
with a correspondingly enhanced respect for embodied information. 
There is some understanding of the algorithmic content of 
descriptions. But so far we have merely scratched the surface.

\section{PHYSICS AND
DEFINABILITY}

 When a Nobel Prize winner in Physics is quoted as 
 saying\footnote{David Gross, quoted in New Scientist, Dec. 10 2005, ``Nobel Laureate 
 Admits String Theory Is In Trouble".}: 
 \begin{quote}
The state of physics today is like it was when we were mystified by 
 radioactivity \dots  They were missing something absolutely fundamental. We 
 are missing perhaps something as profound as they were back then.
 \end{quote}
 people take notice. And this from 2004 winner David Gross did cause 
 something of a stir.

This section is in the nature of a road test for the conceptual framework we have 
been building up around the notion of unconventionality of a computation. We briefly 
outline various gaps in the `standard model' of physics and point to the how a more basic 
viewpoint can help. The discussion will consist of a brief commentary 
centred around some revealing quotations from physicists themselves.

We start with Einstein himself complaining about the resort to ad hoc elements   
of physical theories:
\begin{quote}
\dots I would like to state a
theorem which at present can not be based 
upon anything more than upon a faith in the simplicity, i.e. intelligibility, 
of nature \dots nature is so constituted that it is possible logically to lay down 
such strongly determined laws that within these laws only rationally completely 
determined constants occur (not constants, therefore, whose numerical value 
could be changed without destroying the theory) \dots
\end{quote}
Notice that this is not just an exhortation to physicists to look for a better theory. 
It is an expression of faith in the fact that a theory which successfully 
defines the observable universe should itself be determined by the 
universe. That is, what we observe is there because the universe 
is `self organising' itself, as one would expect of an emergent 
system with sufficient invariance of its structure to exhibit 
a high degree of mathematical rigidity. An interesting question 
is the extent to which constants of the model which make it work, 
but which are not measurable, are actually defined. 
In general, one can interpret the necessity of certain 
values of the constants to make the model work as 
a sort of invariance. What we suspect of invariance is 
an elusiveness of algorithmic infrastructure to the relationship between 
the local and global which makes it possible that aspects of 
reality are dependent on basic information in a way that is 
impossible for us to theoretically unravel. We 
identify below a mathematical model within which to 
host basic computable causality. Characterising 
the automorphisms of this model promises to be a 
key task.  

Here is a more recent questioning of progress towards a 
more comprehensive model of physics, from Peter Woit, author 
of the  book {\em Not Even Wrong -- The Failure of String Theory 
and the Continuing Challenge to Unify the Laws of Physics} (Jonathan Cape, 2006): 
\begin{quote}
By 1973,
physicists had in place what was to
become a fantastically successful theory of fundamental particles 
and their interactions, a theory that was soon to acquire the name of 
the Ôstandard modelÕ. Since that time, the overwhelming triumph of the 
standard model has been matched by a similarly overwhelming failure to 
find any way to make further progress on fundamental questions.
\end{quote} 
And one of Peter Woit's concerns is those undefined constants:
\begin{quote}
One way of thinking
about what is unsatisfactory about the standard model is that it leaves 
seventeen non-trivial numbers still to be explained, \dots
\end{quote}

Alan Guth, originator of the inflationary hypothesis, would like to 
see the laws of physics defined:
\begin{quote}
If the creation of the
universe can be described as a quantum process,
we would be left with one deep mystery of existence: 
What is it that determined the laws of physics?
\end{quote}
If we  think  we are observing the universe  
defining its own laws, we can but hope to 
have access to the defining process in the course of time. 

We are talking here about hugely unconventional 
computation. It may be so unconventional that for 
us it is hardly computation at all. But its existence 
can be framed as a something feasibly 
approachable, at least in principle. Roger 
Penrose\footnote{Roger Penrose: Quantum physics and conscious thought, in 
{\em Quantum Implications: Essays in honour of David Bohm} (B.J. Hiley and F.D. Peat, eds.), pp.106-107.}
 calls it  `Strong Determinism':
 \begin{quote}
 [According to Strong Determinism] \  \dots\  all the complication, variety and 
 apparent randomness that we see all about us, as well as 
 the precise physical laws, are all exact and unambiguous 
 consequences of one single coherent mathematical structure.
 \end{quote}
 
 In our final section, we fill in the missing ingredient --- namely, the 
 fundamental mathematical host for all this embodied information, 
 definability and higher order computation. Before that, 
 a remark regarding mathematical structures: Mathematical 
 structures commonly consist of objects connected  by 
 operations or relations. Sometimes the difference 
 between these classes is blurred, but in an interesting structure 
 there are objects which accumulate {\em information} 
 expressive of their context in the structure.  Sometimes this 
 information can be `read' by the relations on the  structure, 
 which express a formal `causality', 
whereby the distribution of information itself has a structure. 
 This appears to be a feature of our own universe. \enlargethispage{12pt}

\section{MODELLING BASIC CAUSALITY}

Another quotation, this time from Lee Smolin's 2006 book on {\em 
The Trouble with Physics}, p.241:
\begin{quote}
\dots causality itself is fundamental \dots 
\end{quote}
The `early champions' of the role of causality mentioned by Smolin -- 
Roger Penrose, Rafael Sorkin, Fay Dowker,  Fotini Markopoulou -- 
make a doughty bunch, formidable protagonists in contemporary 
 turf wars around quantum gravity, causal sets and 
a hydra-headed superstring theory. The aim, as outlined by 
Smolin, is a more 
comprehensively immanent 
universe\footnote{Lee Smolin, {\em The Trouble With Physics}, p.241.}:
\begin{quote}
It is not only the case that the spacetime geometry determines 
what the causal relations are. This can be turned around: Causal relations 
can determine the spacetime geometry \dots 
ItÕs easy to talk about space or spacetime emerging from something more 
fundamental, but those who have tried to develop the idea have found 
it difficult to realize in practice. \dots  We now believe they failed because they 
ignored the role that causality plays in spacetime. These days, many of us 
working on quantum gravity believe that causality itself is fundamental -- 
and is thus meaningful even at a level where the notion of space has disappeared.
\end{quote}

Note that we are talking about very specific observed and computationally 
well-served {\em causality} here, which largely frees us from the strictures 
of John Earman\footnote{In  {\em A Primer On Determinism}, 
D. Reidel/Kluwer, 1986, p.5.} 
regarding the more wide-ranging use of the term:

\begin{quote}
\dots  the most venerable of
all the philosophical definitions
[of determinism] holds that the world is deterministic just in case 
every event has a cause. The most immediate objection to this approach 
is that it seeks to explain a vague concept - determinism - in terms of a truly 
obscure one - causation.
\end{quote}
In fact, a primary objective of this modelling of basic computable causality 
is the clarity that the mathematics of the model brings to less 
easily described causality, and to issues regarding over causation, 
downward causation and non-locality.

The question is, what kind of causality fails to engage with the informational content 
of the reality it structures? The relevance of the question derives from the fact 
that it is causal structure from which information derives and whereby it is 
stored. Computation is about information, and potentially equipped to 
model the way in which the basic causality of our universe respects and transports 
information. Once again it was Turing who gave us a precise formulation 
corresponding to the fundamentality of the intuition. 

Alan Turing's 1939 paper is a neglected masterpiece ---  less cited than the more 
famous trio of papers that gave us the universal Turing machine, the Turing 
test for intelligence, and the mathematics of morphogenesis --- but 
crackling with ideas and perceptive intuitions. The {\em oracle Turing 
machine} as it came to be called appears on just one page of this densely 
argued article. Essentially, it equips the computer -- in the form of a Turing machine -- 
to roam the scientific universe of real numbers, accepting type 1 inputs, and 
outputting, if we are lucky, type 1 outputs. This sometimes described 
as {\em relative} computation. 

An oracle Turing machine exactly expresses the character of basic 
causality in the world, progressively sampling information and 
transferring it comprehensively across time and space. 

The mathematician or computer scientist --- and maybe Turing himself at the time 
of its invention  --- 
regards the oracle machine as a model of how we might compute 
using data given to us from an unknown source. This viewpoint, together 
with observation of the apparent actuality of incomputability in the natural 
universe, provides the basis for Jack Copeland's notion of 
{\em hyper computation} (beyond the Turing barrier). 

But  the physicist is presented with a model --- 
the {\em Turing universe} ---  within which the computable 
content of Newtonian dynamics comfortably fits; at a basic level of course. 
As Poincar\'e speculated, and researchers from Kreisel in 1970, to 
Beggs, Costa and Tucker today observed, more broadly 
interactive contexts based on Newton's laws can generate 
infinitary mathematics with attendant incomputabilities. 

Mathematically, the type-2 computable functions $\Phi$  over the reals  are termed 
{\em Turing} -- or {\em partial computable} (p.c.) -- {\em functionals}. 
Turing, despite his longterm interest in interactive computation 
(mainly between humans and machines), seems to have never 
mentioned his oracle machines again. It was left to another 
highly creative but under-appreciated mathematician, 
Emil Post, 15 years older than Turing, to set in motion 
the mathematical development of Turing's model. 
In 1944 Post defined the {\em degrees of unsolvability} -- later 
called the {\em Turing degrees}  -- as a classification of reals 
in terms of their {\em relative} computability.

Strangely, the subsequent investigation of the mathematical character of 
the Turing degree structure was a process entirely detached from 
reality. There was no sense at all of relevance to the real world. 
The fact that the Turing universe underpinned a wide range of 
dynamical contexts in which the `design' was understood meant 
nothing. The possibility that all sorts of higher structure might 
be better understood via an analysis of  definability or 
invariance in the basic underlying model was never entertained. 
I was there through a golden age of technical 
development. Turing was gone, taking with him his broadly questioning 
brilliance, leaving behind a universally adopted computational paradigm. We 
recursion theorists were busy doing our sums 
while the natural world around us computed in mysterious and 
wondrous ways. There was no such thing as 
unconventional computation. 

There were mathematical events beneath which one can retrospectively 
detect a sort of subliminal prescience. In 1965 Hartley Rogers gave a 
fascinating talk (judging by the 1967 paper\footnote{H. Rogers, Jr.,  
  Some problems of definability in recursive function theory, in    {\em  Sets,
Models and Recursion Theory}   (J. N. Crossley,  ed.), Proceedings of the
Summer School in Mathematical Logic and  Tenth Logic Colloquium, Leicester,
August--September, 1965,  North Holland, Amsterdam,  pp. 183--201.}   
that came out of it) at the 
Tenth Logic Colloquium in Leicester, England. What was 
remarkable was the focus\footnote{Hartley Rogers' 1965 Agenda, in 
'Logic Colloquium '98' (S.R. Buss, P. Hajek and P. Pudlak, eds.), Proceedings of 
the Annual European Summer Meeting of the ASL, held in Prague, Czech Republic, 
August 9-15, 1998, Lecture Notes in Logic 13, Association for Symbolic Logic/A.K. Peters, 
Natick, Massachusetts, pp. 154-172.} 
on the large scale structure of the Turing 
universe, via the notions of invariance and definability 
that we have identified as relevant to the  emergence of 
form in wide range of different environments. There was 
in evidence a `Hartley Rogers Agenda', built around a number 
of deep and difficult questions about the global character 
of the Turing degrees. Over the years, there has been 
a growing intuition that 
Rogers' questions are key to pinning down how higher order relations 
on the real world can appear to be computed. 
Much of the progress with these questions rests on 
the  richness of Turing structure discovered so far. 
Mathematically structural pathology is a disappointment. 
Out in the real world pathology is super-abundant, 
both generator of and avatar of a  richness of real world definability. 
In the Turing universe. the pathology takes on a parallel role, 
 becoming the raw material for a multitude of definable  relations, 
 counterparts to  visibly `computable' structure out in the real world. 

\section{UNDEFINABLE RELATIONS}

We might have developed the view of unconventional computation as 
higher type computation  in various guises -- emergence, definability etc -- in other 
contexts. Another fruitful workspace would have been that of 
artificial intelligence/neuroscience. Each has it's special strengths. 
That of the physics is the clear way in which it displays the 
fragility of definability, and  the consequences of its 
failure. 

Failures of definability are not necessarily negative in their impact. 
Many friendly features of our universe depend on them. In physics, 
there is a wide range of special symmetries underpinning aspects 
of our observed world. A symmetry is of course an instance of 
an automorphism at work, maybe small-scale or very selective 
in its scope. More important broad impact symmetries 
include the relationship of the SU(3) symmetry group to the 
quark model underlying hadrons, for which Murray Gell-Mann got 
a nobel prize in 1969. One of the interests of such particular 
examples is that they point to the possibility of a 
rich automorphism structure underlying the basic causal 
structure, and hence to the identification on new relations 
defined/computed within the physics with potentially far-reaching  
explanatory power.

Back in the underlying Turing model, there is some disagreement 
about the potential character of the automorphism group of both 
the local and the global structures. An interest in the so-called 
`Bi-interpretability Conjecture' originating with Leo 
Harrington goes back around 30 years, during which time 
various people have managed to prove partial versions 
of the conjecture, with interesting consequences for the 
automorphism groups. Essentially, what the conjecture says 
is that there is a close enough correspondence between 
the structure of the Turing degrees and that of second order 
arithmetic for the two structures to share a number 
of characteristics, particularly related to automorphisms and 
definability. A full verification of bi-interpretability would 
impose rigidity on the Turing universe, and 
invalidate it as a model for the real universe, which appears 
to be far from rigid. There is no consensus of informed guesses 
concerning rigidity. 

Failure of rigidity would have potentially dramatic consequences 
for the longstanding search for a `realistic' interpretation 
of quantum non locality and the collapse of the wave function 
in conjunction with a measurement. What we commonly have is the 
deterministic continuous evolution of the wave equation describing 
a physical system via Schr\"odinger's equation, involving 
the superposition of basis states. We may then have a 
probabilistic non-local discontinuous change due to 
a measurement  -- and 
observe a jump to a single basis state. There are various 
interpretations of this. The simplest is that what we 
are encountering is a level of failure of definability 
at the ontological level of the quantum world --- there 
is just not enough connectivity and information down there to 
uniquely identify basis states. While the intervention 
entailed by the measurement changes the situation. 
If it is a higher type relationship of definability gets 
unconventionally computed, it is allowed to 
operate non-locally, without any of the usual problems 
for the physics. \enlargethispage{-22pt}

There are wider ramifications to such a `realistic' and 
immanent deciding of physical transitions. The Many Worlds 
interpretation and its Multiverse derivatives begin to look 
pleasingly redundant. It is not good that this becomes yet another 
misfortune to impact on  Hugh Everett III and his 
family --  nowadays represented by his son Mark, the talented 
lead singer of the EELS. No doubt, even with such a powerfully 
persuasive replacement for Many Worlds and 
the various Multiverses, via  unconventional computation, it will 
be hard to divert David Deutsch from his 
view\footnote{David Deutsch, {\em The Fabric of Reality}, Allen Lane, 1997, p.48.} 
that:
\begin{quote} \dots 
understanding the multiverse is a precondition for understanding 
reality as best we can. Nor is this said in a spirit of grim determination 
to seek the truth no matter how unpalatable it may be  \dots  
It is, on the contrary, because the resulting world-view is so much 
more integrated, and makes more sense in so many ways, than 
any previous world-view, and certainly more than the cynical 
pragmatism which too often nowadays serves as surrogate for a 
world-view amongst scientists.
\end{quote}
But there are many others, like presumably George 
Ellis,\footnote{The Unique Nature of Cosmology, in {\em Revisiting the 
Foundations of Relativistic Physics} (eds. Abhay Ashtekar et al), Kluwer, 1996, p.198.} 
would breath a sigh of relief:
\begin{quote} The issue of what is to be regarded as an ensemble of 
`all possible' universes is unclear, it can be manipulated to produce any 
result you want \dots The argument that this infinite ensemble actually exists 
can be claimed to have a certain explanatory economy 
(Tegmark 1993), although others would claim that Occam's razor has been 
completely abandoned in favour of a profligate excess of existential 
multiplicity, extravagantly hypothesized in order to explain 
the one universe that we do know exists.
\end{quote}

There are many other ways in which the admission of 
an extended computational repertoire can bolster the integrity of 
the observed `one universe that we do know exists'. 
Physics is just one area can benefit from the mathematics of 
definability, invariance, emergence and higher type computation. 
Alan Turing would have been fascinated.


\end{document}